# Tunable third harmonic generation in the vacuum ultraviolet region using dielectric nanomembranes

Kuniaki Konishi[1,2], Daisuke Akai[3], Yoshio Mita[4], Makoto Ishida[3], Junji Yumoto[1,5], Makoto Kuwata-Gonokami[5]

[1]Institute for Photon Science and Technology, The University of Tokyo, Tokyo, 113-0033, Japan
[2]PRESTO, Japan Science and Technology Agency, Saitama 332-0012, Japan.
[3]Electronics-Inspired Interdisciplinary Research Institute, Toyohashi University of Technology, Toyohashi, 441-8580, Japan
[4]Department of Electrical and Electronic Engineering, The University of Tokyo, Tokyo, 113-0033 Japan
[5]Department of Physics, The University of Tokyo, Tokyo, 113-0033, Japan

**Abstract**
**Tunable coherent light sources operating in the vacuum ultraviolet (VUV) region in 100–200-nm (6–12 eV) wavelength range have important spectroscopic applications in many research fields, including time-resolved angle-resolved photoemission spectroscopy (ARPES). Recent advances in laser technology have enabled the upconversion of visible femtosecond lasers to the vacuum and extreme ultraviolet regions. However, the complexity of their experimental setups and the scarcity of bulk nonlinear crystals for VUV generation have hampered its widespread use. Here, we propose the use of a free-standing dielectric nanomembranes as a simple and practical method for tunable VUV generation. We demonstrate that third harmonic VUV light is generated with sufficient intensity for spectroscopic applications from commercially available $SiO_2$ nanomemebranes of submicron thicknesses under excitation with visible femtosecond laser pulses. The submicron thickness of the nanomembranes is optimal for maximize the VUV generation efficiency and prevents self-phase modulation and spectral broadening of the fundamental beam. The observed VUV photons are up to $10^7$ photons per pulse at 157 nm with 1-kHz repetition rate, corresponding to a conversion efficiency of $10^{-6}$. Moreover, the central VUV wavelength can be tuned in 146–190-nm wavelength range by changing the fundamental wavelength. We also explore material and thickness dependence with experiments and calculations. The presented results suggest that dielectric nanomembranes can be used as a practical nonlinear media for VUV spectroscopic applications.**

## 1. Introduction

Vacuum ultraviolet (VUV) coherent light sources have been a powerful spectroscopic probe [1], allowing the observation of the electronic states of excited atoms [2] and molecules [3]. In the field of life sciences, electronic circular dichroism (ECD) is a powerful tool because of its sensitivity to structural conformation of biomolecules [4]. ECD measurements require efficient sources of VUV light [5] but have already shown promise for probing important biomolecules [6, 7]. Another application that has recently emerged is the time-resolved angle-resolved photoemission spectroscopy (ARPES) [8–11]. By combining ARPES with pulsed lasers in a conventional pump-probe setup, it is possible to directly explore the dynamics of non-equilibrium electronic states. In ARPES, high-energy photons must overcome the work function of solids, typically approximately 5 eV. VUV photons have energies well above this value; this facilitates the effective probing of a large area of the momentum space within the Brillouin zone [12]. Moreover, the use of high-energy photons increases bulk sensitivity owing to its increased propagation length in the material [13]. This is particularly important for investigating the macroscopic quantum phenomena, such as superconductivity [9]. Furthermore, as output increases, VUV light is expected to be useful for controlling chemical reactions [14] and accordingly, for use in lithography [15] and laser processing [16].

Despite its numerous potential uses, coherent VUV light is still challenging to generate. Currently, third harmonic generation (THG) [17] and higher harmonic generation (HHG) [18] from gases are demonstrated. However, these generation schemes are often complicated and require specialized techniques for stable operation. For its usage to increase, realizing a more practical and convenient solid-state-based method for wavelength conversion into the VUV region is essential. A problem here is that the conventional bulk nonlinear crystals, such as β-BaB$_2$O$_4$, LiB$_3$O$_5$, and CsLiB$_6$O$_{10}$, cannot be used for generating VUV pulses owing to their strong absorption. An exception is the KBe$_2$BO$_3$F$_2$ (KBBF) crystal [19, 20]; however, KBBF crystals are difficult to grow and are unavailable [21]. Moreover, in these previous methods, it has been demanded to satisfy the phase matching condition to increase the maximum average power of generated harmonic wave [22]. This requirement has limited the choice of materials that can be used as the nonlinear medium.

Recent technological advances have drastically improved the performance of femtosecond lasers. In particular, the use of solid-state laser diodes for pumping Yb-doped media significantly increases the stability and reduces the maintenance cost associated with regenerative amplifiers [23, 24]. Using these femtosecond lasers for generating second and third harmonics paves the way toward the development of new practical tabletop sources of coherent VUV light. In this case, high conversion efficiency can be expected owing to the high electric field strength of the femtosecond laser that removes the restriction that the phase matching condition must be satisfied and dramatically expands the choice of nonlinear media. In fact, recently, new types of solid nonlinear materials for VUV generation, such as surfaces of transparent solids in the visible region [25, 26] and dielectric metasurfaces [27, 28], which are artificial nanostructures with sizes comparable with or smaller than the light wavelength, have been proposed and demonstrated.

In these methods where the nonlinear medium does not satisfy the phase matching condition, the thickness of the nonlinear medium that contributes to VUV generation should be the same order of the coherent length which is typically several-hindered nm. It is because, even if the thickness of the nonlinear medium is larger than the coherent length, it does not contribute to increase the harmonic intensity [22]. In this case, $F_{max}(n\omega)$, which is the maximum photon flux of the $n$-th order harmonic wave can be generated by the femtosecond laser light source with the repetition rate of $N$, is simply described as

$$F_{max}(n\omega) = \frac{I_{max}(n\omega)}{\hbar n\omega} \propto N\left\{\alpha\left[E_{max}(\omega)\right]^n\right\}^2, \tag{1}$$

where $I_{max}(n\omega)$ is the maximum average power of generated $n$-th order harmonic wave, $\alpha$ is the $n$-th order nonlinear coefficient, and $E_{max}(\omega)$ is the maximum electric field amplitude of the fundamental beam. To increase $F_{max}(n\omega)$, the value of $N$ has been increased using an oscillator with a high MHz-order repetition rate [25, 26] and the value of $\alpha$ has been effectively enhanced using the metasurfaces owing to the effect of Mie resonances in the dielectric nanostructures [27, 28]. Despite these efforts, the power of the VUV light generated from these methods are still too weak to be used for practical application, such as laser ARPES and ECD spectroscopy. Moreover, these methods relinquish advantage of phase-matching independence for their power gain, thus losing the ability of simple wavelength tunability; it is difficult to change the oscillator wavelength over a wide wavelength range, and the resonance wavelength of a metasurface is uniquely determined by its shape.

Another way to increase the generated VUV power, as apparent from equation (1), is by increasing the maximum electric field amplitude of fundamental beam $E_{max}(\omega)$. This approach is particularly important for applications where the maximum value of the repetition rate of light source $N$ is fixed, such as time-of-flight-type ARPES [29]. However, in the case of femtosecond laser pulse excitation, $E_{max}(\omega)$ is limited by two major factors: the first is the laser damage threshold of solid-state materials; the second is self-phase modulation incurred during propagation through a bulk substrate material. Such nonlinear propagation effects significantly reduce the maximum electric field amplitude of the fundamental beam and the efficiency of wavelength conversion when femtosecond pulses with energies from micro- to milli-Joules are used. In the method of the previous research [25-27], although the only a very thin region near the surface contribute to VUV generation, the fundamental wave propagates through a bulk substrate before reaching the nonlinear medium and the effects of nonlinear propagation there are inevitable.

To overcome these problems for increasing the fundamental pulse intensity, in this study, we propose and demonstrate the use of dielectric free-standing nanomembranes with a thickness of several-hundred nanometers or less as a new solid-state material for VUV THG. The submicron thickness of the nanomembranes is optimal for maximize the VUV generation efficiency and prevents self-phase modulation and spectral broadening of the fundamental beam. In addition, the high damage threshold of the dielectric facilitates an increase in the excitation power of the fundamental laser pulse. We demonstrate that THG in $SiO_2$ nanomembranes excited at the fundamental 470-nm wavelength with commercially available femtosecond optical parametric amplifiers (OPAs) enables coherent VUV light generation at 157-nm wavelength with sufficient photon flux (~$10^7$ photons per pulse with 1-kHz repetition rate; $10^{10}$ photon per second) to be used as a probe beam for VUV spectroscopy, including laser ARPES. Furthermore, the resonance-free nature of our structure combined with the OPA light source allows continuous tuning of the wavelength in the VUV region (146–190 nm) by simply changing the wavelength of the fundamental beam. We also investigate the dependence of THG intensity on the excitation power as well as the laser damage threshold for materials of different thicknesses. We show that $SiO_2$ can achieve the maximum photon number because of its large damage threshold, whereas epitaxial γ-$Al_2O_3$ enables high generation efficiency. As the repetition rate is increased to 100 kHz, the VUV THG intensity linearly increased with it. Owing to these excellent properties and because these dielectric nanomembranes are commercially available and easy to handle, they are promising and practical nonlinear media for wavelength conversion to the VUV region for spectroscopic applications.

## 2. Estimation of the VUV THG intensity generated from a dielectric nanomembrane

First, we estimate the VUV THG intensity generated from a dielectric nanomembrane. When the depletion of the fundamental wave is negligible and the nonlinear material is isotropic, the intensity of the third harmonics (TH) beam, $I_3$, can be expressed using the following equation [22]:

$$I_3 = \frac{(2\pi)^2 |\chi_{1111}|^2}{16 n_1^3 n_3 \varepsilon_0^2 c^2 \lambda_3^2} I_1^3 L^2 \mathrm{sinc}^2 \frac{|3k_1 - k_3|L}{2}, \qquad (2)$$

where $I_1$ is the intensity of the fundamental beam, $\chi_{1111}$ is the third-order nonlinear susceptibility, $\lambda_3$ is the wavelength of the TH beam, $L$ is the thickness of the nonlinear media, and $n_1$ and $n_3$ are refractive indices of the fundamental and TH beams, respectively. $k_1 = 2\pi n_1/\lambda_1$ and $k_3 = 2\pi n_3/\lambda_3$ represent wavenumbers for the fundamental and TH beams, respectively. $\varepsilon_0$ (~$8.85 \times 10^{-12}$ C/m·V) is the dielectric constant of vacuum, and $c$ (~$3.00 \times 10^8$ m/s) is the speed of light. That is, for a 300-nm-thick $SiO_2$ nanomembrane ($\chi_{1111}$ = 2.8 × $10^{-22}$ $m^2/V^2$ [22], $\lambda_3$ = 1.57 × $10^{-7}$ m, $n_1$ = 1.47 and $n_3$ =1.66 [30], $|3k_1 - k_3|$ = 7.48 × $10^7$ $m^{-1}$) at a fundamental wavelength of 470 nm (THG is 157 nm) and intensity $I_1$ of $10^{17}$ W/$m^2$, the THG conversion efficiency is as high as $I_3/I_1 = 4.2 \times 10^{-6}$. It is worth noting that for a pulse duration of 100 fs, the chosen intensity of the fundamental beam corresponds to a fluence of 1 J/$cm^2$, which is just below the damage threshold [31]. The estimated conversion efficiency indicates that the nanomembrane enables VUV pulse energy that is sufficient for VUV spectroscopy, including laser ARPES. Because third-order susceptibility is of the order of $10^{-22}$ $m^2/V^2$ [22] for most dielectrics, similar conversion efficiency can be achieved using other dielectric membranes.

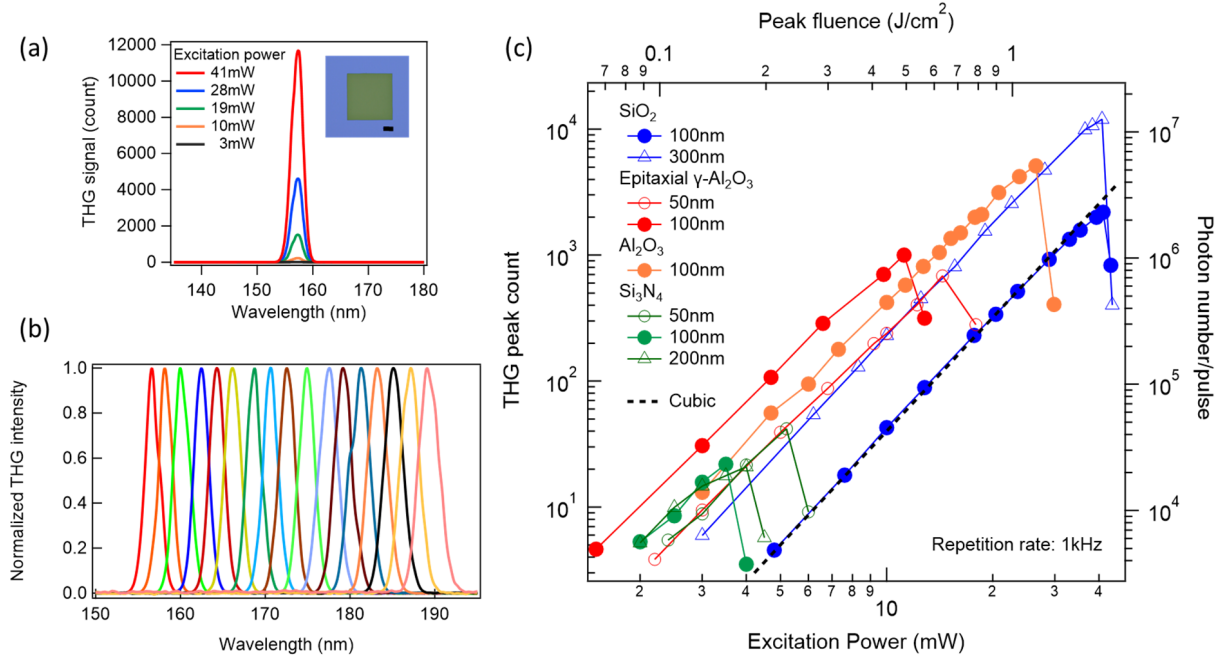

Fig. 1. Vacuum ultraviolet third harmonic generation from dielectric nanomembranes with a 1-kHz repetition rate excitation (a) Dependence of VUV THG spectra from a 300-nm-thick $SiO_2$ nanomembrane on the excitation power. All spectra presented in Fig. 1 were acquired with a 0.5-s integration time. The inset is a microcopy image of the sample. Length of scale bar is 100 μm. (b) Dependence of VUV THG spectra on the excitation wavelength, which changes from 470 to 572 nm at 6-nm intervals. Each spectrum is normalized by the peak intensity. (c) Dependence of VUV THG intensity from dielectric membrane samples ($SiO_2$ (blue), epitaxial γ-$Al_2O_3$ (red), $Al_2O_3$ (orange), and $Si_3N_4$ (green)) on the excitation power. Peak fluence at the focal point is shown in the top axis, and VUV photon number par pulse is shown in the right axis. The dashed line indicates cubic dependency.

## 3. Measurement of VUV THG from dielectric nanomembranes

To demonstrate that the above conversion efficiency can be achieved in practice, we performed THG experiment using commercially available, free-standing $SiO_2$ nanomembranes (100- and 300-nm thick), $Al_2O_3$ (100-nm thick), $Si_3N_4$ (50-, 100-, and 200-nm thick), and self-made epitaxial γ-$Al_2O_3$ (50- and 100-nm thick) (see Supplementary Material Section 1). As a typical example, the microscope image of a 300-nm-thick $SiO_2$ nanomembrane is shown in the inset of Fig 1(a). The epitaxial γ-$Al_2O_3$ was grown on a (100) silicon substrate through chemical vapor deposition [32], whereas the silicon substrate was removed using Deep-RIE to produce a free-standing thin film (see Supplementary Material Section 1). In the epitaxial γ-$Al_2O_3$, the crystal axis perpendicular to the substrate is arraigned but the in-plane crystal axis is randomly oriented, the typical domain size of which is approximately several tens nm [33]. Therefore, the optical response is isotropic at normal incidence. Stress control is usually important for obtaining flat membranes: when compressive stress is introduced in a film, which is often the case for Si and thermally oxidized $SiO_2$, the film becomes wrinkled [34, 35]. Here, however, no buckling was observed in the epitaxial γ-$Al_2O_3$; thus, a flat membrane could be formed without additional stress-control processes.

For the VUV THG experiment, we used an OPA with a pulse duration of approximately 100 fs at a repetition rate of 1 kHz, pumped by a regeneratively amplified femtosecond Ti:Sapphire laser (see Supplementary Material Section 2). The VUV THG spectra of a 300-nm-thick $SiO_2$ membrane excited with a wavelength of 470 nm are presented in Fig. 1 (a). The figure presents clear VUV THG signals, which increase with the excitation power, at a wavelength of 157 nm (7.9 eV). Figure 1(b) presents the observed VUV THG spectrum of a 300-nm $SiO_2$ membrane with a peak value when the wavelength of the fundamental beam output from the OPA is changed from 470 to 572 nm at 6 nm intervals while keeping the excitation power constant at 10 mW. Each spectrum is normalized by the maximum intensity. We can see that the generation of VUV coherent light from the $SiO_2$ nanomembrane is possible at any wavelength greater than 157 nm, without significantly degrading the signal to noise ratio. We also confirmed that the generation of coherent VUV light from the $SiO_2$ nanomembrane is possible at as short as approximately 146 nm (see Supplementary information Section 3). The shortest wavelength of VUV THG is believed to be determined by the wavelength at which VUV light reabsorption by $SiO_2$, where it occurs here at the bandgap of $SiO_2$, at approximately 7.5 eV [36].

Figure 1(c) presents the excitation power dependence of the THG peak intensity for each nanomembrane when a 157-nm (7.9 eV) THG is generated using a 470-nm (2.6 eV) fundamental beam. It can be observed that, for all nanomembranes, the THG intensity is proportional to the cubic of the excitation power and that the THG signal sharply decreases as soon as the excitation power exceeds the irreversible damage threshold [31]. For wide-bandgap materials, such as $Al_2O_3$ and $SiO_2$, it is known that the density of free electrons (which are generated through multiphoton or tunneling ionization) increases exponentially due to the avalanche processes. Material damage occurs when it reaches a critical density [37]. We did not observe a deviation from the cubic dependence of the THG intensity on the fundamental pulse power until the damage occurred.

Figure 1(c) also shows that the peak damage threshold fluences for $Si_3N_4$, $Al_2O_3$, and $SiO_2$ nanomembranes are approximately 0.2, 1.2, 1.8 J/$cm^2$, respectively. Because the bandgap energies of $Si_3N_4$, $Al_2O_3$, and $SiO_2$ are 4.8 [38], 6.5 [36], and 7.5 eV [36], respectively, one may conclude that the larger the bandgap, the higher the damage threshold. This is consistent with previous results obtained for bulk materials [38] and membranes [31]. Moreover, the inter-band transition in $Si_3N_4$ is a two-photon process, whereas the transition from the valence to conduction band in $Al_2O_3$ and $SiO_2$ requires three photons. That is, the damage threshold is significantly increased in $Al_2O_3$ and $SiO_2$ because their free-electron generation rate is much lower than that in $Si_3N_4$. Though the bandgap of epitaxial γ-$Al_2O_3$ has not been experimentally determined yet, it was found that its damage threshold is about 0.5 J/$cm^2$. Figure 1(c) also shows that the damage threshold of the membranes is almost independent of their thicknesses. This is because for subwavelength-thick membranes, the electric field strength at the back surface is maximized due to interference [39], i.e., damage is expected to occur there independent of the thickness.

It is worth noting that the dependence of the THG signal with the thickness varies with the material of the nanomembrane. In the $Si_3N_4$ nanomembranes, the 50-nm-thick sample yields a lower TH wave intensity than the 100-nm-thick sample but a higher intensity than the 200-nm-thick sample. In contrast, in the $SiO_2$ nanomembranes, the THG intensity increases approximately by 10 times as the thickness increases from 100 to 300 nm. This is because the intensity of the TH beam is determined by the absorption length of the sample rather than the film thickness, as we will discuss later.

| Sample (thickness) | Maximum photon number of VUV THG (photon/pulse) |
|---|---|
| $SiO_2$ (300 nm) | $1.4 \times 10^7$ |
| $SiO_2$ (100 nm) | $2.6 \times 10^6$ |
| Epitaxial $\gamma$-$Al_2O_3$ (100 nm) | $1.1 \times 10^6$ |
| Epitaxial $\gamma$-$Al_2O_3$ (50 nm) | $8.0 \times 10^5$ |
| $Al_2O_3$ | $5.9 \times 10^6$ |
| $Si_3N_4$ (200 nm) | $2.4 \times 10^4$ |
| $Si_3N_4$ (100 nm) | $2.6 \times 10^4$ |
| $Si_3N_4$ (50 nm) | $5.7 \times 10^4$ |

Table I. Maximum photon number of VUV THG from nanomembranes

In our experiments with 1-kHz repetition rate and approximately 50-μm beam spot size, the maximum VUV photon flux was achieved by the 300-nm $SiO_2$ nanomembrane when it was excited with a 41-mW fundamental beam power, as shown in Fig. 1(c). This is because the laser damage threshold of $SiO_2$ is higher than that of other nanomembranes, and further, the THG intensity in the $SiO_2$ nanomembrane increases as the film thickness increases. It is worth noting that the epitaxial $\gamma$-$Al_2O_3$ membranes show the highest VUV THG efficiency below the damage threshold.

We used a photomultiplier as a detector to quantitatively estimate the number of VUV photons generated from the 300-nm $SiO_2$ membrane (see Supplementary Material Sections 2 and 4). The VUV THG signal intensity observed at an excitation wavelength of 470 nm and power of 10 mW was 2881 nVs (the data is shown in Supplementary Material Section 4), which approximately corresponds to $2.6 \times 10^5$ photon/pulse. The observed THG counts can be converted to photon numbers using this result, as shown on the right axis of Fig. 1(b). The maximum VUV photon number from the samples can also be estimated and summarized, as presented in Table I. Based on this result, the maximum number of photons generated from the 300-nm-thick $SiO_2$ nanomembrane with a 41-mW excitation can be estimated to be about $1.4 \times 10^7$ photon/pulse and $1.4 \times 10^{10}$ photon/s, with 1-kHz repetition rate, as presented in Fig. 1(b). This corresponds to an average power of about 18.1 nW, which gives a conversion efficiency of about $10^{-6}$. This value is consistent with the estimation results discussed at the beginning of this paper. Note that the observed VUV intensity is more than two orders larger than that reported in previous methods [25, 27, 28].

## 4. Numerical evaluation of VUV THG efficiency dependence on the thickness of nanomembranes

THG efficiency depends strongly on the thickness of the nanomembranes. Figure 2 presents the results of the THG efficiency as a function of nanomembrane thickness for $Si_3N_4$, $SiO_2$, and $Al_2O_3$ [40] (see Supplementary Material Section 6). Here, unlike equation (2), the refractive indexes are treated as complex numbers to consider the effect of absorption in the VUV region. All the plots were normalized using the corresponding maximum THG intensity. For $SiO_2$ and $Al_2O_3$, oscillations with a period of several-hundred nanometers were observed due to phase mismatch. In addition, due to the finite absorption of VUV light, its envelope tends to decrease as the membrane thickness increases. Therefore, in materials with small absorption in the VUV region, the THG efficiency is maximized when the film thickness is about the same as the coherence length (see Supplementary Material Section 6). This corresponds to approximately 400 nm for $SiO_2$ and approximately 200 nm for $Al_2O_3$. In contrast, no oscillation was observed in the $Si_3N_4$ nanomembrane, and its THG intensity was maximum at a thickness of approximately 100 nm, which becomes nearly constant for thicknesses over 200 nm. This is because $Si_3N_4$ has a strong absorption (absorption coefficient of $7.76 \times 10^7$ m$^{-1}$, see Supplementary Material Section 6) as its bandgap, which is less than the photon energy of the generated VUV light, i.e., only VUV photons generated near the back surface are emitted to the outside.

We also observed that the spectrum of the fundamental beam pulse is broadened in the bulk dielectric material due to self-phase modulation and that the THG signal decreased significantly (see Supplementary Material Section 7). At a pulse width of 100 fs or higher, it is difficult to avoid self-phase modulation even if pre-chirping is applied to the fundamental beam through pulse shaping. These results are consistent with the dependence of the THG intensity on thickness for the nanomembranes observed in Fig. 1(c), indicating the importance of using nanomembranes rather than the bulk form for an efficient generation of VUV THG.

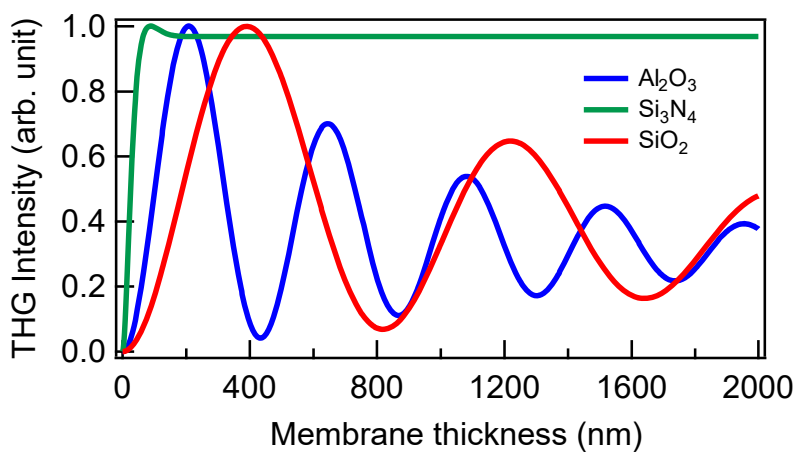

Fig. 2. Numerical simulation of the dependence of THG intensity on material thickness. THG intensity dependence on membrane thickness for different materials (blue: $Al_2O_3$, green: $Si_3N_4$, red: $SiO_2$). Each plot is normalized by the maximum intensity.

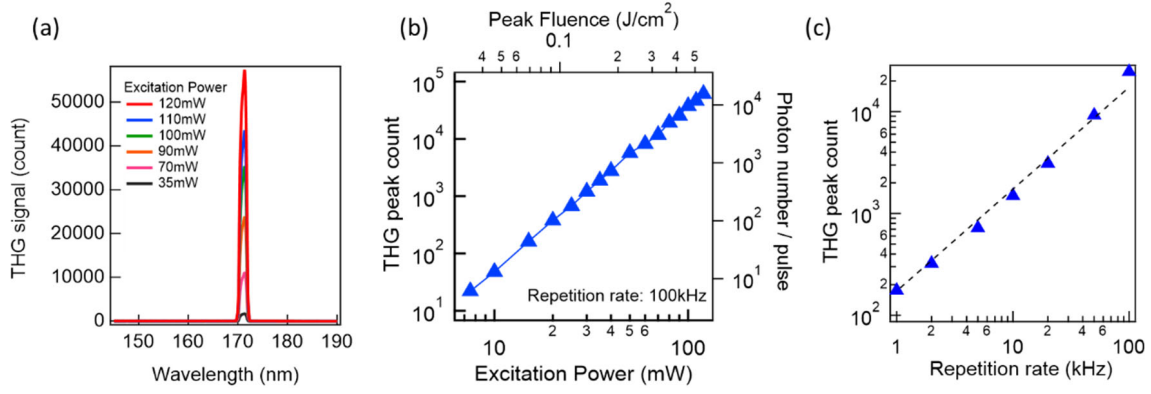

Fig. 3. Results of VUV THG measurement for SiO$_2$ materials with a 100-kHz repetition rate excitation (a) Dependence of VUV THG spectra from a 300-nm-thick SiO$_2$ membrane on the excitation power. Dependence of VUV THG intensity from a 300-nm-thick SiO$_2$ membrane on excitation power (b) and repetition rate (c).

## 5. VUV THG intensity dependence on the repetition rate of the fundamental beam

Because higher repetition rates above 1 kHz are often required in practice, we also examined the VUV THG intensity at a repetition rate of 100 kHz (See Supplementary Material Section 2). The dependence of the obtained THG spectrum and intensity on the fundamental beam intensity is presented in Fig. 3(a) and (b), respectively. The THG intensity also shows the cubic dependence on the intensity of the fundamental beam for a repetition rate of 100 kHz. No damage occurred even at an excitation power of 120 mW because the peak fluence (0.55 J/cm$^2$) did not reach the damage threshold (1.8 J/cm$^2$) observed in Fig. 1(c). The number of VUV photons generated at an excitation intensity of 100 mW was estimated using the photomultiplier, as described above. The observed signal intensity was 191.5 nVs/pulse, which corresponds to approximately $1.0 \times 10^4$ photon/pulse and $1.0 \times 10^9$ photon/s at a repetition rate of 100 kHz. Figure 3 (c) shows that the THG intensity also increases in proportion to the repetition rate.

In the 100-kHz rate experiment, the excitation intensity of 100 mW corresponds to a fluence of approximately 0.46 J/cm$^2$. Because the damage threshold of the SiO$_2$ membrane is 1.8 J/cm$^2$ (see Fig. 1(c)), it is expected that the excitation intensity can be further increased by 3.9 times, that giving 59 times higher THG intensity. Furthermore, if the repetition rate of the fundamental beam is multiplied by 10, i.e., 1 MHz, the nanomembrane turns to a VUV coherent light source capable of generating $5.9 \times 10^5$ photons/pulse and $5.9 \times 10^{10}$ photons/second. Realizing this requires a femtosecond laser with a repetition rate of 1 MHz and average power of 3 W as an excitation laser. Such lasers are commercially available. For time-of-flight-type ARPES suitable for time-resolved ARPES measurement [29], the number of photons per pulse is limited to $10^3$ photons/pulse or less because the number of photons entering the detector must be at most one, and the maximum repetition rate of light sources is around 4 MHz [41], due to the limited response speed of the detector. Thus, the current performance of our membranes is practical for use in typical ARPES setups. Moreover, the results must also be applicable to long-pulse excitation applications, where energy resolution is important. If the pulse width is increased from 100 fs to 1 ps to increase the energy resolution to 2 meV, the peak electric field intensity becomes one-tenth, and correspondingly the THG becomes one thousandth the original strength. If we assume the same 1-MHz repetition rate and 3-W average power system, even in this case, the number of photons is $5.9 \times 10^2$ photons/pulse, which is sufficient for the measurement. For hemispherical-type ARPES, the maximum photon number is also approximately $10^9$ per second, limited by the damage threshold of the detector [41]. Therefore, the above specification is also sufficiently applicable to light sources for hemispherical-type ARPES. The 157-nm VUV observed in Fig. 1 has almost the same wavelength as the VUV emitted from the KBBF crystal [20], so it can be replaced without much difficulty.

## 6. Conclusion

In conclusion, we demonstrated the wavelength conversion to the VUV region using THG in dielectric nanomembranes. We showed that a 300-nm $SiO_2$ nanomembrane excited by femtosecond pulses radiated from an OPA pumped by a regenerative amplifier can generate 7.9-eV (157 nm) photons at a 1-kHz repetition rate, with $1.4 \times 10^7$ photon/pulse, which corresponds to 18.1 nW and the conversion efficiency of as much as $10^{-6}$. Furthermore, wavelength conversion can be performed in a wide wavelength range (146–190 nm) by simply changing the wavelength of the fundamental wave. Our numerical simulation shows that coherent VUV light can be generated with the highest efficiency at 400-nm thickness of $SiO_2$ and 200-nm thickness of $Al_2O_3$, which is consistent with the experimental results. We also demonstrated VUV generation at 100 kHz and confirmed the generation of $1.0 \times 10^4$ photon/pulse, which is sufficient for laser ARPES applications.

The dielectric free-standing nanomembrane used in this experiment are all commercially available and inexpensive, excluding γ-$Al_2O_3$ nanomembrane which we created. In addition, it can generate coherent VUV light simply by placing them in the optical path of the excitation beam. Therefore, by combining a commercially available regenerative amplifier and an OPA system, practical tunable coherent VUV light source can be configured, which can be applicable for VUV spectroscopic applications, for example, to laser ARPES.

As future prospects, the fabrication of nanostructures on nanomembranes has the potential to realize advanced optical control techniques based on structural effects, such as photonic crystals and metamaterials [42–44]. While this loses the advantage of wavelength tunablilty, the high damage threshold should allow for continued application to high-peak-intensity pulses. This furthermore allows for the effective use of the solid-state nature of our material, which is difficult with gas-based generation schemes. Combined with nanostructure fabrication, nanomembranes may become a powerful and versatile new platform for VUV upconversion from high-pulse-energy lasers.

**Supplementary Materials**

See supplementary material for the details of samples, VUV THG measurements and their results and the evaluation of generated VUV THG intensity and photon numbers.

**Acknowledgments**

We thank Y. Svirko, H. Sakurai, Y. Arashida, D. Hirano, T. Ikemachi, and N. Kanda for their helpful discussion and E. Lebrasseur, M. Fujiwara, and A. Mizushima for their support in fabricating the device.

Fabrication of the samples was performed using the apparatus at the VLSI Design and Education Center (VDEC) of the University of Tokyo. This research was supported by JST PRESTO (JPMJPR17G2), JSPS KAKENHI (18H01147), MEXT Q-LEAP(JPMXS0118067246), MEXT Photon Frontier Network Program, MEXT "Nanotechnology Platform" (12024046), and the Center of Innovation Program funded by the Japan Science and Technology Agency.

**Data Availability Statements**

The data that supports the findings of this study are available within the article and its supplementary material.

# Supplementary Materials for

# Tunable third harmonic generation in the vacuum ultraviolet region using dielectric nanomembranes

Kuniaki Konishi[1,2]*, Daisuke Akai[3], Yoshio Mita[4], Makoto Ishida[3],

Junji Yumoto[1,5] and Makoto Kuwata-Gonokami[5]**

[1]Institute for Photon Science and Technology, The University of Tokyo, Tokyo, 113-0033, Japan

[2]PRESTO, Japan Science and Technology Agency, Saitama 332-0012, Japan.

[3]Electronics-Inspired Interdisciplinary Research Institute, Toyohashi University of Technology, Toyohashi, 441-8580, Japan

[4]Department of Electrical and Electronic Engineering, The University of Tokyo, Tokyo, 113-0033 Japan

[5]Department of Physics, The University of Tokyo, Tokyo, 113-0033, Japan

1. **Samples**

    The total areas of the $SiO_2$ nanomembrane, $Si_3N_4$ nanomembrane, and $Al_2O_3$ nanomembrane were 500 × 500 μm$^2$, 1 × 1 mm$^2$, and 300 × 300 μm$^2$, respectively. These nanomembranes are commercially available (the $SiO_2$ and $Si_3N_4$ nanomembranes were purchased from TEMwindows.com, and the $Al_2O_3$ nanomembrane was purchased from Lebow Company).

    The following method is used for the preparation of epitaxial γ-$Al_2O_3$ nanomembrane. First, 48-nm-thick γ-$Al_2O_3$ thin film was epitaxially grown on a 525-μm-thick Si (100) substrate through chemical vapor deposition. A Si(100) substrate was cleaned by ammonium hydrogen-peroxide mixture ($NH_4OH:H_2O_2:H_2O$ = 0.05:1:5), hydrochloric hydrogenperoxide mixture ($HCl:H_2O_2:H_2O$ = 1:1:6), and diluted hydrofluoric acid ($HF:H_2O$ = 1:50) to remove organic contaminants, particles, metallic contaminants, and thin oxide film on the Si substrate. Then the γ-$Al_2O_3$ film was grown via metal organic chemical vapor deposition at 960 °C and 500 Pa with trimethylaluminum and $O_2$ gas sources [30, S1]. γ-$Al_2O_3$ films were characterized using reflection high energy electron diffraction. A photoresist was spin-coated at 3000 rpm for 60 s to protect the epitaxial γ-$Al_2O_3$ thin film, which is then baked for 1 min at 120 °C. Subsequently, the back of the Si substrate is spin-coated with an electron beam (EB) resist OEBR-CAP112 (1000 rpm, 60 s) and baked for 1 min at 120 °C. 300 × 300 μm$^2$ square holes were patterned using a rapid EB lithography system (F7000S-VD02, Advantest) and baked for 1.5 min at 120 °C and developed in 2.38% tetramethyl ammonium hydroxide (TMAH). The silicon substrate under the membrane structure was removed using the Deep-RIE process (MUC-21 ASE-Pegasus, SPP technologies). Here, to avoid damaging the epitaxial γ-$Al_2O_3$ thin film, the etching is stopped while leaving a few microns of silicon. Then, the sample is placed in acetone to remove photoresist from the front side. Finally, it was dipped in 2.38% TMAH at 80 °C for approximately 1 h to remove the remaining silicon, after which it is placed in 2-propanol (IPA) for cleaning. This method was followed to fabricate a nanomembrane with a maximum size of 1 × 1 mm$^2$.

## 2. VUV THG measurement

A schematic of the experimental setup is presented in Fig. S1(a). The light source was an optical parametric amplifier (Light Conversion Ltd., HE-TOPAS) with a pulse duration of approximately 100 fs at a repetition rate of 1 kHz. It was pumped by a regeneratively amplified femtosecond Ti:Sapphire laser (Coherent Inc., Astrella). Linearly polarized laser pulses whose power was controlled by a neutral density filter were introduced into the vacuum chamber (approximately $7.0 \times 10^{-1}$ Pa) through a quartz window and focused on the samples using a plano-convex lens (f = 300 mm) at a normal incidence with a diameter of approximately 50 μm (Fig. S1(b)). The beam profile on the focal point was measured directly using a CCD beam profiler. The fundamental beam was stopped using two VUV bandpass filters (eSource optics, 25150FNB for 157 nm and 25170FNB for 171 nm) placed behind the samples. The THG signals collinear to the incident beam were incident on a VUV spectrometer (McPherson, Model 234/302) with a cooled charge-coupled device (CCD) camera (ANDOR Technology, DO950P BR-DD) or a photomultiplier tube (PMT, Hamamatsu Photonics K.K, R8486). The grating groove density was 1200 g/mm. The photomultiplier was used with an applied voltage of 500 V. The VUV signal is usually detected using the CCD camera but can also be detected using the PMT by setting a mirror inside the spectrometer. The measured spectra obtained using the CCD camera can be observed on a PC connected by it. The current signal output from the PMT was amplified using an amplifier unit (Hamamatsu Photonics K.K, C12419) and measured using an oscilloscope (IWATSU, DS-5654A).

For the 100-kHz repetition rate experiment, we used second harmonic wave (514 nm) generated by using a 1-mm-thick BBO crystal pumped by a Yb:KGW laser (Light Conversion Ltd., PHAROS) as a fundamental beam. It has a center wavelength of 1028 μm and a pulse duration of 190 fs, operated at a repetition rate of 100 kHz. The excitation pulse is focused on the 300-nm $SiO_2$ membrane using f = 100 lens with a diameter of approximately 20 μm to observe the 171-nm VUV THG (Fig. S1(c)).

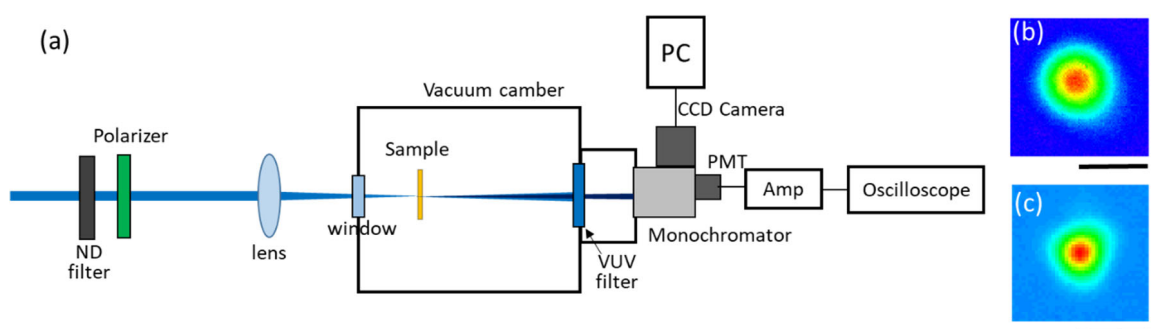

Fig. S1. (a) Schematic diagram of the experimental setup (b)(c) Beam profiles of the fundamental beam on the sample position ((b): 1-kHz repetition rate (the scale bar is 50 μm), (c): 100-kHz repetition rate (the scale bar is 20 μm)).

## 3. Wavelength tunability of VUV THG

To clarify the shortest THG wavelength generated from the 300-nm $SiO_2$ membrane, we also used an OPA (Light Conversion Ltd., ORPHEUS-HP) with a pulse duration of approximately 190 fs at a repetition rate of 6 kHz. It was pumped by a regeneratively amplified femtosecond Yb:KGW laser (Light Conversion Ltd., PHAROS) with a center wavelength of 1028 μm and a pulse duration of 190 fs operated at a repetition rate of 6 kHz. The excitation pulse was focused on a 300-nm $SiO_2$ membrane using f = 100 lens.

Figure S2 presents the observed VUV THG spectrum for a 300-nm $SiO_2$ membrane, showing the peak value as the wavelength of the fundamental beam output from the OPA is changed from 470 to 434 nm at 6 nm intervals while keeping the excitation power constant at 10 mW. It can be seen that the minimum wavelength of THG is approximately 146 nm, excited by 434-nm fundamental beam.

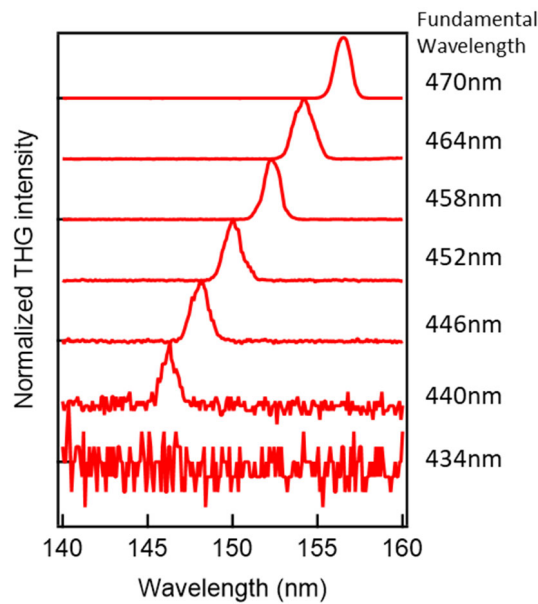

Fig. S2. Dependence of VUV THG spectra on the excitation wavelength, which changes from 470 to 434 nm in 6 nm steps, as indicated on the right side of the figure. Each spectrum is normalized by its peak intensity. All spectra were acquired at 0.5-s integration time.

## 4. Output THG signal from the photomultiplier observed using an oscilloscope

To quantitatively estimate the number of VUV photons generated from the 300-nm $SiO_2$ membrane, we performed an experiment using a photomultiplier as a detector (see Supplementary Material Section 1). The observed VUV THG signal for an excitation wavelength of 470 nm and power of 10 mW is presented in Fig. S3. Its integrated intensity is 2881 nVs.

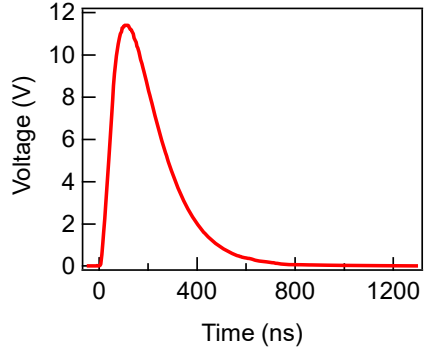

Fig. S3. VUV THG signal obtained using the photomultiplier for an excitation wavelength of 470 nm and power of 10 mW. It is an integration result for 8024 pulses.

5. **Evaluation of VUV photon number**

The relation between the number of photons ($x$) emitted from the sample and magnitude of the voltage signal ($S$ [V·s]) observed using oscilloscope can be expressed as follows:

$$x = S \cdot \frac{n}{F \cdot D \cdot R \cdot g \cdot \eta \cdot G},$$
(S1)

where F represents the transmittance of the VUV filter, D is the diffraction efficiency in the grating in the VUV spectrometer, R is the reflectivity of the VUV mirror to the PMT port, n is the number of electrons in a total charge of 1C, g is the gain of the PMT, $\eta$ is the quantum efficiency of the PMT, and G [V/A] represents the gain of the amplifier unit after PTM.
Their values are summarized in Table S1.

Table S1

|  | 157 nm | 171 nm |
|---|---|---|
| Transmittance of VUV filter ($F$) | $(0.15)^2$ | $(0.20)^2$ |
| Grating diffraction efficiency ($D$) | 0.32 | 0.30 |
| Reflectivity of VUV mirror ($R$) | 0.74 | 0.74 |
| Number of electrons of 1C ($n$) | $6.242 \times 10^{18}$ | |
| Gain of PMT ($g$) | $10^5$ | |
| Quantum efficiency of PMT ($\eta$) | 0.13 | |
| Gain of amplifier unit ($G$) [V/A] | $10^6$ | |

6. **Numerical simulation of the VUV THG intensity**

The THG intensity $I^{3\omega}$ generated from the membrane excited at normal incidence is described as follows [38]:

$$I^{3\omega} = \frac{64\pi^4}{c^2} \left| \frac{\chi^{(3)}}{\varepsilon_\omega - \varepsilon_{3\omega}} \right| (I^\omega)^3 \left| e^{i\psi^\omega} T \left(1 - e^{-i\Delta\psi}\right) \right|^2,$$
(S2)

where $I^\omega$ represents the intensity of the fundamental beam; $\chi^{(3)}$ is the third-order nonlinear susceptibility; c is the speed of light; $\varepsilon_\omega$ and $\varepsilon_{3\omega}$ are the permittivity of the fundamental and third harmonic waves, respectively; and $\Delta\psi$ is the phase angle, which is given by the following:

$$\Delta\psi = \psi^\omega - \psi^{3\omega} = \frac{6\pi L}{\lambda_\omega}\left(n^\omega - n^{3\omega}\right), \quad (S3)$$

where L is the thickness of the membranes, $\lambda_\omega$ is the fundamental wavelength, and $n^\omega, n^{3\omega}$ are the fundamental and THG beam refractive indices. T represents the transmission factor independent on L (see Ref. 38).

In this case, the dependence of $I^{3\omega}$ on L is simply described as follows:

$$I^{3\omega} \propto \left|\left(1 - e^{-i\Delta\psi}\right)\right|^2. \quad (S4)$$

The refractive indices of the membranes used in the calculation are presented in Table S2. Absorption coefficient α can be calculated using $\alpha = 4\pi\kappa/\lambda$, where $\kappa$ is the imaginary part of the refractive index. In this calculation, we assumed that the intensity of the fundamental wave does not change due to THG. Coherence length $L_c$ is described as follows [38]:

$$L_c = \frac{\lambda_\omega}{6\left|n^\omega - n^{3\omega}\right|}. \quad (S5)$$

When L equals $L_c$, $\Delta\psi$ equals π. The values of $L_c$ calculated from Table S2 are 413 and 216 nm for $SiO_2$ and $Al_2O_3$, respectively.

Table S2

| Material | Wavelength (nm) | Refractive index | Reference |
|---|---|---|---|
| $SiO_2$ | 470 | 1.47 | [28] |
| | 157 | 1.66 + i0.0185 | |
| $Al_2O_3$ | 470 | 1.69 | [S2] |
| | 157 | 2.05 + i0.0262 | [S3] |
| $Si_3N_4$ | 470 | 2.05 | [S4] |
| | 157 | 2.64 + i0.97 | |

## 7. VUV THG from a thick SiO2 sample

The VUV THG spectra from a 500-μm-thick bulk fused silica generated through excitation with a source with a wavelength of 470 nm under the same conditions used to obtain Fig. 1(a) are presented in Fig. S4 (see Supplementary Material Section 1 for the details of the experiment). Compared with Fig. 1(a), it can be seen that the THG spectra from bulk fused silica have broadened and that the intensity is much less than that from the membrane.

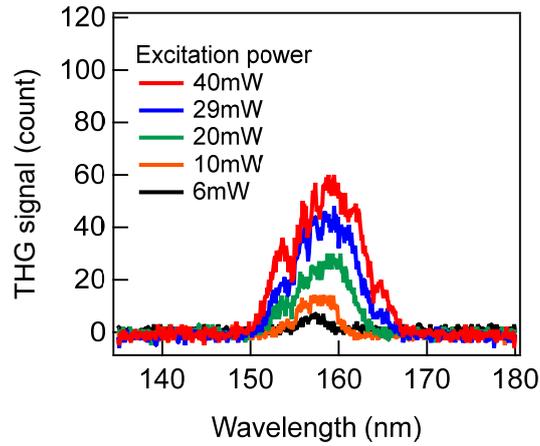

Fig. S4. Dependence of VUV THG spectra from a 500-μm-thick $SiO_2$ membrane on the excitation power.

When the thickness of the nonlinear medium is large, the spectrum of the fundamental wave spreads as the excitation pulse energy increases due to the self-phase modulation effect. However, the peak intensity of the wavelength-converted beam does not change. The self-phase modulation effect is the major factor that reduces the efficiency of VUV THG, together with absorption in the VUV region.